# Numerical optimization of a nanophotonic cavity by machine learning for near-unity photon indistinguishability at room temperature

Guimbao Joaquin*[1], Sanchis Lorenzo[1], Weituschat Lukas[1], Llorens J M[1], Song Meiting[2], Cardenas Jaime[2], Postigo Pablo[1,2]

[1]Instituto de Micro y Nanotecnología, IMN-CNM, CSIC (CEI UAM+CSIC), Tres Cantos, Madrid, E-28760 Spain
[2] The Institute of Optics, University of Rochester, Rochester, New York 14627, USA
*j.guimbao@csic.es

**Abstract:** Room-temperature (RT), on-chip deterministic generation of indistinguishable photons coupled to photonic integrated circuits is key for quantum photonic applications. Nevertheless, high indistinguishability ($I$) at RT is difficult to obtain due to the intrinsic dephasing of most deterministic single-photon sources (SPS). Here we present a numerical demonstration of the design and optimization of a hybrid slot-Bragg nanophotonic cavity that achieves theoretical near-unity $I$ and high coupling efficiency ($\beta$) at RT for a variety of single-photon emitters. Our numerical simulations predict modal volumes in the order of $10^{-3}(\lambda/2n)^3$, allowing for strong coupling of quantum photonic emitters that can be heterogeneously integrated. We show that high $I$ and $\beta$ should be possible by fine-tuning the quality factor ($Q$) depending on the intrinsic properties of the single-photon emitter. Furthermore, we perform a machine learning optimization based on the combination of a deep neural network and a genetic algorithm (GA) to further decrease the modal volume by almost three times while relaxing the tight dimensions of the slot width required for strong coupling. The optimized device has a slot width of 20 nm. The design requires fabrication resolution in the limit of the current state-of-the-art technology. Also, the condition for high $I$ and $\beta$ requires a positioning accuracy of the quantum emitter at the nanometer level. Although the proposal is not a scalable technology, it can be suitable for experimental demonstration of single photon operation.

## Introduction

Indistinguishable single photons are the leading candidates for quantum communication and quantum information processing technologies. They play a central role in a range of proposed schemes, including quantum simulation[1], quantum walks[2], boson sampling[3], quantum teleportation[4], and quantum networks[5]. However, the complex mesoscopic environment of solid-state sources entails fundamental barriers that restrict the operation to cryogenic temperature ($T$)[6]. Trying to overcome the thermal restrictions of quantum devices remains a challenge for the development of on-chip, on-demand SPS. A feasible approach for achieving efficient indistinguishable photon emission from a solid-state emitter consists of maximizing the emitter-field coupling ($g$) through the effective confinement of light in an ultra-small cavity mode-volume ($V_{eff}$) and reaching the strong coupling regime[7]. In this regime the transfer rate between the emitter and the cavity field exceeds the dephasing rate of the emitter, and the emitted photons are able to leave the cavity before being affected by decoherence[7]. Plasmonic cavities with sub-nanometer gaps between dimers like Au spheres[8], Ag nanowires[9], and surface plasmon-polariton systems[10] or metallic bowties with CdSe/ZnS quantum dots[11] produce the highest $g$ value up to 200 meV [11] and the lowest quality factors ($Q\sim10$)[9]. There are different proposals to improve $Q$ and $\beta$ in these systems, some of them involving dielectric-core/metal-shell schemes for $Q$[10,12], or hybrid FP-nanoantenna cavities for $\beta$[13,14]. However, using plasmonic cavities faces two obstacles[15]: (i) the placement of the emitter in the point with the strongest cavity field can be challenging; (ii) ohmic and quenching losses can be very high. The use of dielectric cavities can avoid the latter limitation and strong coupling can happen using strategies to decrease the modal volume, like slotted photonic crystals. Discrete slotted nanobeams[16,17] lead to volumes in the order of $10^{-3}(\lambda/2n)^3$ while keeping high $Q$. However, because introducing a finite slot causes a large perturbation to the optical mode, $\beta$ values remain low. Continuous-slot designs improve $\beta$ and $Q$[18], and more recently, slot-anti slot concatenations in 1D-PC[19,20] have shown record $Q/V_{eff}$ ratios with PC cavities. Also, designs based on cascaded cavities schemes have shown promising results with dielectric structures[21]. According to those works, a slotted dielectric cavity can provide sufficient small modal volume for strong coupling, thus a high $I$, avoiding at the same time the losses inherent to plasmonic cavities. However, for highly dissipative emitters, the dependence of $I$ with $g$ at RT is highly non-trivial[7]. With high $g$ there is a high population transfer rate between the emitter and the cavity field, so the emitted photons must leave the cavity before getting dephased by the emitter. This can be accomplished by setting the right $Q$. As we will show, this trade-off between different rates (i.e. dephasing rate, $g$ and $Q$) translates in to a complex dependence of $I$ with the cavity figures of merit.

In this work, we show that achieving high $I$ at RT requires a tuning of $Q$ together with a small modal volume. That does not translate to a high $Q$, but a specific $Q$ threshold depending on the emitter's intrinsic properties and the modal volume. From our calculations, none of the previously mentioned dielectric cavities can provide a high $I$ for strong dissipative emitters despite achieving small modal volumes. Furthermore, the implementation of machine learning algorithms for the geometrical optimization of the cavity modal volume and $Q$ has shown promising results in recent works[22-25]. Here we present a numerical demonstration of a design strategy for high indistinguishable SPS at RT strongly coupled to a hybrid slot-Bragg waveguide cavity. We vary the geometrical parameters of the waveguide cavity (i.e., the waveguide width, slot width, number of periods), and we obtain a theoretical estimation of the cavity performance for $I$, $\beta$, and the Purcell enhancement. We explore different types of promising SPS (InGaAs[26] and GaAs[27] quantum dots, single molecules[28], localized excitons in transition metal dichalcogenides TMDC monolayers[29], and diamond color centers[30]), and we obtain theoretical near-unity $I$ and high $\beta$ simultaneously by parameter optimization. Finally, we develop a hybrid deep neural network-GA scheme that further reduces the modal volume for achieving near-unity $I$ with a slot width of 20 nm. The optimized device presents strong challenges for current fabrication and quantum emitter (QE) positioning techniques. In this regard, we have developed a comparison of the design requirements with the state of the art demonstrations.

**Methods**

We can compute the value of $I$ for a QE with radiative decay rate $\gamma$ and pure dephasing rate $\gamma^*$ coupled to a photonic cavity (with decay rate $\kappa$ and electromagnetic coupling constant $g$) from the Lindblad equation and applying the quantum non-regression theorem. For each ($g$, $\kappa$, $\gamma$, $\gamma^*$) we have[7]:

$$I = \frac{\iint_0^\infty dt d\tau |<\hat{a}^\dagger(t+\tau)\hat{a}(t)>|^2}{\iint_0^\infty dt d\tau <\hat{a}^\dagger(t)\hat{a}(t)><\hat{a}^\dagger(t+\tau)\hat{a}(t+\tau)>} \quad (1)$$

Where $\hat{a}^\dagger, \hat{a}$ are the creation/annihilation operators of the cavity mode. Details of the calculation can be found in the supplementary material. The values of $g$ and $\kappa$ are linked to $Q$ and $V_{eff}$ by $\kappa \sim 1/Q$ and $g \sim 1/\sqrt{V_{eff}}$.

Figure 1e shows the value of $I$ for photons emitted by a high dissipative QE with $\gamma^* = 10^4 \gamma$ as a function of $g$ and $\kappa$ normalized to $\gamma$ in the coherent strong-coupling regime (i.e. $g > \gamma^* + \gamma$). In this regime, the rate of photon transfer from the emitter to the cavity is $R = 4g^2/\kappa$ [7], which exceeds the pure dephasing rate ($R > \gamma^*$) for certain values of $\kappa$. For high $I$ the photon must escape out of the cavity before the emitter dephases it. In other words, $\kappa > \gamma^*$, which means that a small $Q$ is needed. Specifically, for a QE with $\gamma^*=10^4\gamma$ one needs a value of $\kappa/\gamma$ above $2\cdot 10^4$ for $I>0.9$. The region of high $I$ in Figure 1e has a shape and area that depend on $T$ through $\gamma^*$. For a QE at RT, $\gamma^* \sim 10^4\gamma$ [7] and the minimum value of $g/\gamma$ to achieve $I>0.9$ is $(g/\gamma)_{min} \sim 10^4$. As $\gamma^*/\gamma$ decreases the area of high $I$ grows and $(g/\gamma)_{min}$ decreases.

Figure 1g shows the contour maps of the region with high $I$ ($I>0.9$) as $\gamma^*$ changes. For moderate dissipative emitters ($\gamma^* \sim 10^2\gamma$), the minimum $g/\gamma$ necessary for $I>0.9$ is $(g/\gamma)_{min}=10^3$. As $\gamma^*$ increases $(g/\gamma)_{min}$ grows monotonously, reaching $10^4$ for $\gamma^* \sim 10^4\gamma$. Similarly, the minimum $(\kappa/\gamma)_{min}$ increases from $10^3$ for $\gamma^* \sim 10^2$, to $2\cdot 10^4$ for $\gamma^* \sim 10^4\gamma$. We can use this colormap to plot the cavities mentioned before, according to its performance for $I$. Plasmonic cavities[8-10] can achieve $I>0.9$ even for high dissipative emitters with $\gamma^* \sim 10^4\gamma$. On the other hand, slotted dielectric cavities[16-18] can achieve $I>0.9$ for emitters with $\gamma^*$ between $\sim 10^2\gamma$ to $\sim 2\cdot 10^2\gamma$ and slot-anti slot concatenations in 1D-PC[19] for emitters with $\gamma^* \sim 2\cdot 10^2\gamma$ to $\gamma^* \sim 4\cdot 10^2\gamma$. The cavity shown in[20] is the only one, in the group of dielectric structures, that can reach $I>0.9$ when $\gamma^* > 2\cdot 10^3\gamma$. According to our calculations those dielectric cavities can potentially achieve the region with $I>0.9$ for high dissipative emitters (i.e., QE at RT) just by increasing its cavity decay rate $\kappa$ (i.e., deteriorating its quality factor $Q$). Figure 1b shows the dependence of the value $(g/\gamma)_{min}$ with $T$ for $I>0.9$, calculated for quantum dots of GaAs[31] and InAs[32], organic molecules[33,34] and defects in 2D materials[35]. The evolution of $(g/\gamma)_{min}$ with $T$ shows a proportional increase with a different trend that depends on $\gamma^*$. We can obtain the $(g/\gamma)_{min}$ needed for $I>0.9$ for a QE at an specific $T$ from Figure 1g. It is interesting to observe that for the technologically relevant $T$ of liquid nitrogen (77 K) the same value $(g/\gamma)_{min}=490$ works for InAs and GaAs QDs and 2D materials.

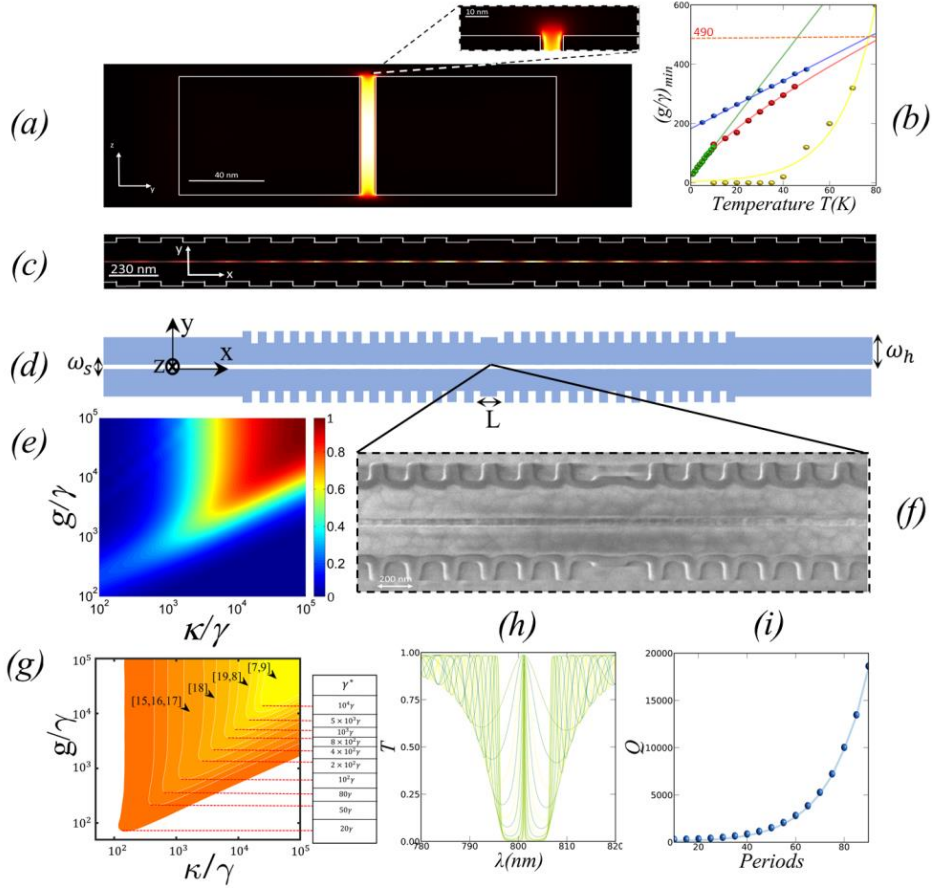

Figure 1. (a) $|E|^2$ field profile in the y-z plane. (b) Variation of the ratio $(g/\gamma)_{min}$ with T for $I>0.9$ and different SPS: GaAs (red), S.molecules (green), 2D-materials (blue), InAs (yellow). (c) $|E|^2$ field profile of the cavity-mode in the x-y plane. (d) Layout of the proposed structure, where $\omega_h$ is the width of each waveguide, $\omega_s$ is the slot width, L is the cavity length, and $\Lambda$ is the grating period. (e) Color map of I as a function of $g/\gamma$ and $\kappa/\gamma$ for photons emitted by a high dissipative QE with $\gamma^* = 10^4 \gamma$. (f) SEM image of the center of the cavity. (g) Contour map of regions with $I>0.9$ for different dephasing values ($\gamma^* = 20\gamma, 50\gamma, 80\gamma, 10^2\gamma, 2\cdot 10^2\gamma, 4\cdot 10^2\gamma, 8\cdot 10^2\gamma, 10^3\gamma, 5\cdot 10^3\gamma$, and $10^4\gamma$). (h) Transmission spectrum of the structure for different number of periods, the FWHM of the resonance scales exponentially with #p. (i) Q versus number of periods.

Therefore, our goal is to keep the $\kappa/g$ ratio inside the region with high I by increasing g and adjusting Q. Moreover, we look for an on-chip cavity that can be CMOS-compatible with photonic integrated circuits (PICs) used in silicon photonics. Slotted one-dimensional dielectric photonic crystal cavities[16-20] have been shown to fulfill most of our requirements in terms of compatibility and small modal volume. Nevertheless, to efficiently control Q, we choose a hybrid slot-Bragg cavity, where Q changes by the number of periods of the Bragg reflector section. Figure 1d shows a layout of our hybrid slot-Bragg photonic cavity aiming to achieve near-unity I and high β simultaneously. $\omega_h$ is the width of each waveguide, $\omega_s$ is the slot width, and #p is the number of periods. While this structure has been explored for sensing applications[36,37,38], it has never been proposed for SPS operation, as far as we know, neither calculated its performance in terms of the figures of merit (I, β). It consists of a phase-shifted corrugated Bragg grating situated at the sides of a Si$_3$N$_4$ ($n_1$=2) deposited on top of a SiO$_2$ substrate ($n_2$=1.4). The cavity length $L$ corresponds to the central section between the two periodic regions and matches the wavelength of the zero-order Fabry-Perot mode for the target wavelength $\lambda$. The Si$_3$N$_4$ thickness ($t$) is set for optimum field enhancement at the slot for the target $\lambda$. Each of the periodic regions behaves like a mirror with an effective reflectivity that depends on the number of periods (#$p$), creating a Fabry-Perot structure. The grating period $\Lambda$ matches the central frequency of the photonic bandgap at the target $\lambda$. In order to get information about the physical behavior of the device, we will set first $\lambda$ = 801 nm to perform a general evaluation of the performance. After that, for each type of emitter the geometrical parameters of the device (i.e. $t$, $L$ and $\Lambda$) are set to match the specific emission wavelength $\lambda$: ($\lambda$, $t$, $L$, $\Lambda$) = (915 nm, 900 nm, 263 nm, 263 nm) for InGaAs[26], (916 nm, 900 nm, 263nm, 263 nm) for GaAs[27], (728 nm, 710 nm, 210nm, 210 nm) for TMDC[28], (785 nm, 770 nm, 225 nm, 225 nm) for S.molecules[29] and (685 nm, 680 nm, 195 nm, 195 nm) for diamond color centers[30].Figure 1a show how the slotted cross-section of the cavity enhances the field of the zero-order TE mode in the gap showing an evanescent tail in the top of the waveguide. This field distribution provide advantages related to the coupling of the source when is heterogeneously integrated on top. The cavity provides strong coupling if the slot width is sufficiently small, and it also provides advantages in extraction efficiency ($\beta$) since (i) cavity and output waveguide share the same cross section, so the modes are perfectly matched; (ii) the integration of the QE (for example colloidal QDs) can be done by direct deposition on top of the cavity which avoids interferences by total internal reflection and enhances $\beta$; (iii) the slot mode has the field maxima at the edges of the slot, which matches well with the region of maximum probability of having SPS in 2D materials deposited on top of waveguides[39]. Finally, the cavity modal volumes are in the order of $10^{-3}(\lambda/2n)^3$ along with the whole slot, increasing the probability of having one or several QE strongly coupled to the cavity mode. As a proof of concept, we have fabricated a specific design valid for diamond color center requirements. We selected ($\omega_s$, #$p$) = (38 nm, 50) and added vertical grating couplers to the structure to collect the input and output light beams. Figure 1f shows an SEM image of the cavity fabricated by e-beam lithography and reactive ion etching on a layer of Si$_3$N$_4$ 130 nm-thick deposited on top of a SiO$_2$ layer (1 $\mu$m-thick) by plasma-enhanced chemical vapor deposition (PECVD). The obtained slot width is $\omega_s$ = 54 nm, and the grating period is 204 nm, with less than 5% of the error to the initial design for the grating period and 30% for $\omega_s$. According to our simulations, the wider slot translates into a modal volume increase, $V_{eff} \sim 6 \cdot 10^{-2}(\lambda/2n)^3$, which slightly reduces the indistinguishability to $I$ = 0.81. This issue can be solved by further optimization by machine learning, as we will show later. We can obtain the transmission spectrum $T(\lambda)$ shown in Figure 1h and the field profile of the cavity mode for a set ($\omega_s$, $\omega_h$, #$p$) using a fully vectorial, bi-directional, frequency-domain model for solving Maxwell's equations (3D-FD)[40]. We obtain $Q$ from $T(\lambda)$ by $Q = \frac{\lambda_0}{FWHM}$ and the cavity decay rate from $\kappa = \omega/2Q$. Details of the model appear in the supplementary material. There is a different effective index for each set ($\omega_s$, $\omega_h$, #$p$), so the values of $\Lambda$ and $L$ are changed to keep the cavity resonance at 801 nm. The volume of the cavity-mode $V_{eff}$ is[41]:

$$V_{eff} = \frac{\iiint \varepsilon(\vec{r})|\vec{E}(\vec{r})|^2 d^3\vec{r}}{max\{\varepsilon|\vec{E}(\vec{r})|^2\}} \qquad (2)$$

The value of $g$, when the QE is placed at the maximum cavity field and perfectly matches the polarization is[42]:

$$g = \frac{\mu_{eg}}{\hbar}\sqrt{\frac{\hbar\omega}{2\varepsilon_M V_{eff}}} \quad ; \quad \mu_{eg} = \frac{3\hbar e^2 f}{2m_{eff}\omega} \qquad (3)$$

Where $\mu_{eg}$ is the electric dipole moment of the excitonic transition, $\omega$ is the frecuency of the transition, $e$ is the electron charge, $\varepsilon_M$ is the dielectric constant in the source region, $\hbar$ is the reduced Planck constant, $m_{eff}$ the exciton effective mass, and $f$ the oscillator strength. Once we have $g$ and $\kappa$, we obtain $I$ according to the procedure outlined in Figure 2a. For the computation of the Purcell enhancement ($\Gamma_p$) and the coupling efficiency $\beta$, we perform 3D-Finite Difference Time Domain (3D-FDTD) simulations[40] by placing a dipole point source emitting at 801 nm with position $x_0$, $y_0$ at the center of the slot and $z_0$ 4 nm above the top of the waveguides. We obtain $\Gamma_p$ by integrating the power $P$ emitted by the source and normalizing it to the power inside a homogeneous environment $P_0$[43]. Finally, we calculate $\beta$ by measuring the fraction of light coupled to guided modes at the output waveguide. Details of the simulations appear in the supplementary material.

Our design strategy can be further enhanced using machine-learning techniques, especially to keep critical fabrication parameters, like the slot width $\omega_s$, experimentally accessible and far from too narrow and unrealistic values. Recently, the optimization of nanophotonic structures by deep learning techniques has been reported[22-25]. The two main advantages are: (i) further improved performance beyond the time-consuming method of sweeping the ($\omega_s$, $\omega_h$, #p) parameters; (ii) we can introduce a vast number of new parameters for the optimization, like the width of each of the Bragg corrugations, as shown in Figure 2b.

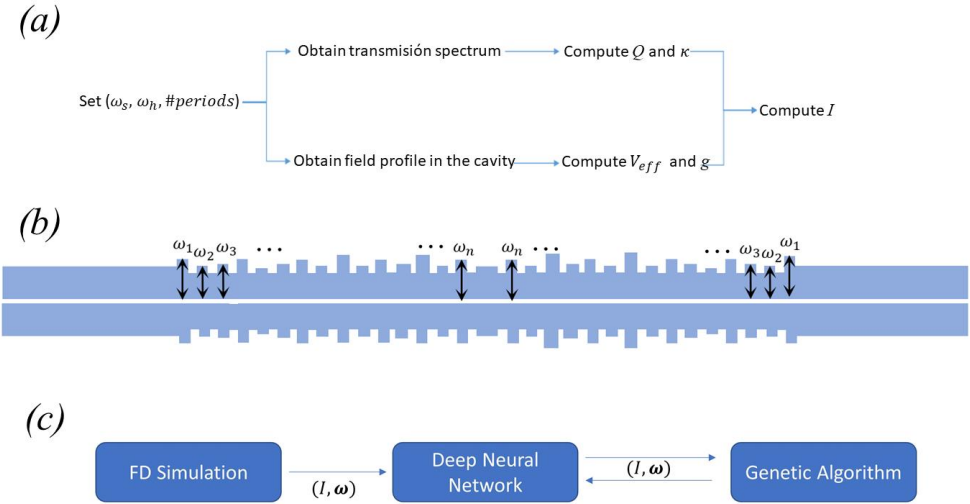

Figure 2 (a) Outline of the computation algorithm for the calculation of *I*. (b) Parametrization of the Bragg corrugations for machine learning optimization. Each $\omega_i$ represents the width of the corresponding Bragg corrugation. (c) Routine for the hybrid NN-GA optimization of the Bragg corrugations.

For that purpose we can use a vector $\boldsymbol{\omega} = (\omega_1, \omega_2, \omega_3, \dots, \omega_n)$, where each entry $\omega_i$ with $i=1,\dots,20$ representing the width of each Bragg corrugation. For each configuration $\boldsymbol{\omega}$ we obtain $I$ using the two-step method described in Figure 2c. We use a GA to create a random vector $\boldsymbol{\omega}$ and the fitness function obtains $I$ from the 3D-FD simulation (Figure 2a). Through the iteration of cross-over and mutation, the GA should find the optimal configuration for maximizing $I$ after a certain number of generations. Details of the code appear in the Supplementary material. However, this procedure faces a critical issue. Typically, in a GA optimization one needs to generate about $10^5$ pairs ($\boldsymbol{\omega}, I$) and the generation of each pair ($\boldsymbol{\omega}, I$) involves a 3D-FD simulation that may take several minutes, making the whole optimization process unfeasible in terms of time and computational resources. To solve this issue, we take a different approach: (i) we generate 5000 pairs ($\boldsymbol{\omega}, I$) through 3D-FD simulations; (ii) with these data, we train a deep neural network (NN) which learns to estimate the outcome of $I$ for any possible $\boldsymbol{\omega}$. Now we can use the NN to calculate $I$ for the fitness function of the GA optimization. In this way, the calculation of the fitness function for each $\boldsymbol{\omega}$ takes just a few seconds; (iii) We perform the GA optimization by calculating the fitness function for each individual of the population through the NN. With this scheme, we reduce by two orders of magnitude the number of actual numerical simulations for the dataset from $10^5$ to $10^3$ with the aid of the NN.

**Results and discussion**

We first assess the performance of the cavity by sweeping the main geometrical parameters and setting a target $\lambda$ = 801 nm. $t, L$ and $\Lambda$ are set to ($t, L, \Lambda$) = (800 nm, 230 nm, 230 nm). Figure 3 shows how $I$ changes with ($\omega_s, \omega_h$) and #$p$ when $\gamma^* = 10^4 \, \gamma$ (a typical ratio for many QE at RT as we have seen before). Figure 3a shows $I$ versus $\omega_h$ and $\omega_s$ for #$p$=10 with $\omega_s$ varying between 10 to 50 nm and $\omega_h$ between 150 to 220 nm (required for single-mode operation). With #$p$ fixed, $Q$ remains constant ($Q$=50), while the field profile of the cavity mode varies for each ($\omega_h, \omega_s$). Therefore, the variation of $I$ follows the variation of $g$ with $\omega_h$ and $\omega_s$. As $\omega_s$ increases, the cavity mode spreads out from the slot and gets confined at each waveguide core separately. That results in an exponential decay of the field intensity in the slot region[44], increasing $V_{eff}$ exponentially with $\omega_s$. Since $g \sim 1/\sqrt{V_{eff}}$, $g$ decreases, driving the system to the weak coupling regime (i.e., going downwards in Figure 1a) and inducing an exponential decay of $I$. For small enough $\omega_s$ (<20 nm) the system remains in the strong coupling regime and $I$ becomes independent of $g$ [7]. Therefore, we can observe that for $\omega_s$ <20 nm $I$ shows a weak variation with $\omega_h$. When $\omega_s$ >20nm, the cavity starts to perform away from the strong coupling regime and $I$ shows an evident change with $\omega_h$, which we will further analyze later. A slot width $\omega_s$ <10 nm produces a maximum value of $I$=0.96, decaying with $\omega_s$ at a rate of $5 \cdot 10^{-3}$ nm$^{-1}$. Figure 3d shows the dependence of $I$ with #$p$, with #$p$ in the range from 10 to 100 and fixed $\omega_h$= 140 nm and $\omega_s$ =15 nm so we keep the strong coupling regime. As #$p$ increases, the effective reflectivity also increases, and the $Q$ factor grows exponentially (see Figure 1i). Consequently, $\kappa$ decreases exponentially with #$p$. Therefore, the time that the photon stays in the cavity increases exponentially with #$p$ and when $\kappa < \gamma^*$ the photon is dephased by the emitter (i.e., going in the left direction in Figure 1a). The result is that $I$ decreases with #$p$ giving $I$ =0.4 for #$p$=100. Figure 3b shows $\Gamma_p$ versus ($\omega_s, \omega_h$) when #$p$=10, $\omega_s$ in the range 10-100 nm and $\omega_h$ between 110 nm to 600 nm. Since $\Gamma_p \sim 1/V_{eff}$, $\Gamma_p$ changes with $\omega_s$ in a similar way than $I$ does. As the slot mode spreads over the waveguide cores, the field´s intensity at the source´s position decreases and $\Gamma_p$ shows an exponential decay. The change with $\omega_h$ displays a more complex structure, shown more clearly in Figure 3e. For $\omega_s$ =15 nm and $\omega_h$ =80, $\Gamma_p$ increases monotonically as the zero-order cosine/even[45] slot mode gets more efficiently confined in the waveguide. $\Gamma_p$ is maximum ($\Gamma_p$ = 11) when $\omega_h$ =125 nm and the strongest light confinement in the waveguide happens. For higher $\omega_h$ the mode spreads over the structure producing a decay of the overlapping with the source that scales with $1/\omega_h$. The decay interrupts abruptly when the zero-order sin-type/odd mode cut-off is reached at $\omega_h$ =155 nm. From there, the same pattern reproduces until the activation of the subsequent mode, and so on. The same behavior happens for $\omega_s$. However, as $\omega_s$ increases the dependence of $\Gamma_p$ with $\omega_h$ shifts to lower values of $\omega_h$. This is because the $\omega_h$ cut-off value of the zero-order sine mode/odd decreases monotonically with $\omega_s$[45]. Therefore, the activation of the second mode shifts to lower values of $\omega_h$ as $\omega_s$ increases.

Figure 3c shows $\beta$ versus $\omega_s$ and $\omega_h$ for the same values of #p, $\omega_s$ and $\omega_h$ used in Figure 3b. While $\Gamma_p$ is a measure of the field enhancement due to the overlapping of all available modes, $\beta$ accounts just for the overlapping with guided modes. Therefore, we expect a similar dependence and, in fact, $\beta$ shows an exponential decay with $\omega_s$ similarly to $I$ and $\Gamma_p$. The dependence with $\omega_h$ shows the same "mode jumps" found for $\Gamma_p$, giving a maximum $\beta$=75% at $\omega_h$ =128 nm. In this case, the regions of high $\beta$ become bigger for higher $\omega_h$, as the number of available modes increases with $\omega_h$.

The position of the QE inside the cavity plays a relevant role[46]. To explore the effect of the position of the QE in $\Gamma_p$ we have performed 3D-FDTD simulations changing the position ($y_0$) of the QE along the cavity cross-section (y-axis) at $z_0$=4 nm above the top of the cavity. Figure 3f shows $\Gamma_p$ versus $y_0$ varying from -225 nm to +225 nm when $\omega_h$ =200 nm, $\omega_s$ =30 nm and #p=10. Since $\Gamma_p$ is proportional to the field of the available modes for each spatial position, the plot reproduces the field profile of the zero-order mode of the slot waveguide. The maximum $\Gamma_p$ happens in the region inside the slot, with maxima at the edges of the waveguides. The enhancement falls abruptly inside the waveguide, with values reduced by one order of magnitude. For a QE located away from the outer edges of the waveguide cores, the evanescent coupling increases the enhancement slightly. In summary, even for a strong dissipative emitter with $\gamma^* = 10^4 \gamma$, we can achieve $I$ >0.9 by adjusting the number of periods and reducing the slot width $\omega_s$ below 10 nm. At the same time, high Purcell enhancement ($\Gamma_p$ =45) and good extraction efficiency ($\beta$=0.7) can be obtained for the same $\omega_s$. On the other hand, we need an accurate positioning of the emitter inside the slot region.

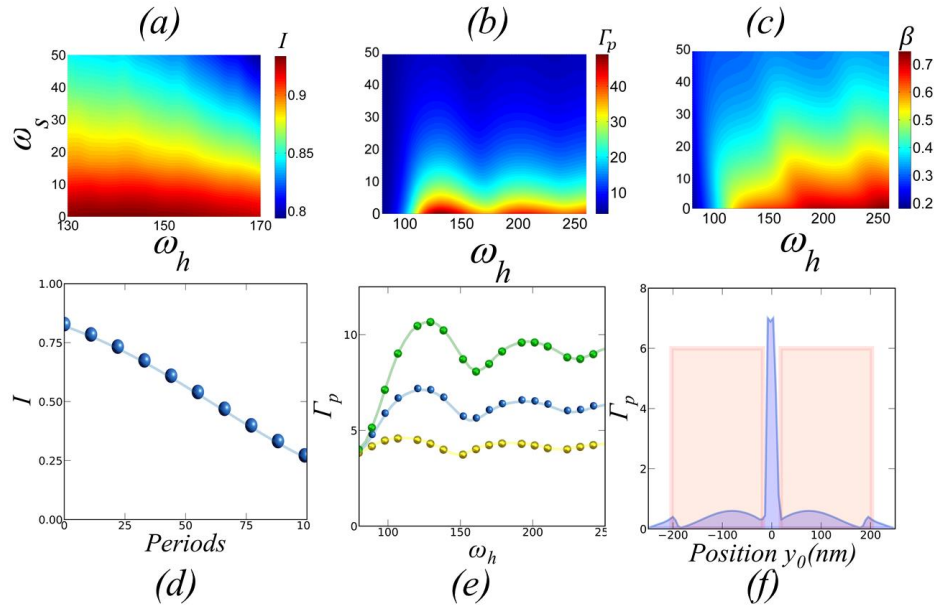

Figure 3. (a) Cavity-induced $I$ when $\gamma/\gamma^* = 10^4$ versus waveguide width ($\omega_h$) and slot width ($\omega_s$) for #p=10. (b) Purcell enhancement ($\Gamma_p$) versus waveguide width ($\omega_h$) and slot width ($\omega_s$). (c) Coupling efficiency ($\beta$) versus waveguide width ($\omega_h$) and slot width ($\omega_s$) for #p=10. (d) $I$ versus number of grating periods (#p) for ($\omega_s$, $\omega_h$) = (5 nm,

140 nm). (e) $\Gamma_p$ versus $\omega_h$ for three $\omega_s$ (green- $\omega_s$ =15 nm, blue- $\omega_s$ =20 nm, yellow- $\omega_s$ =25 nm). (f) $\Gamma$ versus source position $y_0$ along the $y$-axis.

|  | $\gamma^* = 10^2\gamma$ | $\gamma^* = 10^3\gamma$ | $\gamma^* = 10^4\gamma$ |
|---|---|---|---|
| InGaAs | (43,100) | (36,50) | (15,10) |
| GaAs | (41,100) | (30,50) | (9,10) |
| TMDC | (36,120) | (25,60) | (5,12) |
| S.molecules | (40,120) | (28,60) | (8,12) |
| Diamond | (45,100) | (38,50) | (15,10) |

Table 1. Maximum ($\omega_s$ (nm), #p) for $I$>0.9 using InGaAs QD, GaAs QD, TMDCs and single molecules as QE.

We further explore the performance of the device and the design requirements for different types of QE with different dephasing rates. For each type of emitter the geometrical parameters of the device (i.e. $t$, $L$ and $\Lambda$) are set to match the specific emission wavelength $\lambda$. Table 1 shows the values of the pairs ($\omega_s$, #p) needed for $I$>0.9 for five different $\gamma^*/\gamma$ values corresponding to each emitter. The values of the oscillation strengths are extracted from InGaAs[47], GaAs[48], TMDC[49,50], single molecules[28,51] and diamond[52]. We observe that as $\gamma^*$ increases (i.e., $T$ increases) the cavity demands smaller $\omega_s$ (i.e., narrower slot). For the highest oscillator strength (~5 in InGaAs QD and diamond color centers) $(g/\gamma)_{min}$ is easily reached when $\omega_s$ <44 nm and $\gamma^* = 10^2\gamma$. A TMDC QE with oscillator strength ~ 0.1 demands $\omega_s$ <38 nm on the opposite side. In an intermediate situation, the oscillation strength of the GaAs QD (~ 1) gives $\omega_s$ <42 nm. From this, we can find the optimal design for each emitter at high $T$. InGaAs at 300 K have a pure dephasing of 600 $\gamma$ [53], so ($\omega_s$, #p) = (36 nm, 50) are needed for $I$>0.9. GaAs at 300 K has 1450 $\gamma$ [54] and needs of the same values ($\omega_s$, #p) = (36, 50). High dissipative emitters with dephasing of ~$10^4\gamma$ at 300 K, like TMDC[55] and single molecules, demand narrower slot widths ($\omega_s$, #p) = (5 nm, 10). For color centers in diamond, with $\gamma^* = 10^3\gamma$ at room T[56], the optimal configuration is ($\omega_s$, #p) = (38 nm, 50).

As we have shown, for high dissipative emitters with $\gamma^* = 10^4\gamma$ the width of the cavity slot must be $\omega_s$ <10 nm for $I$ >0.9. Similarly, $\omega_s$ <10 nm is needed for $\beta_s$>0.7. At the same time, the emitter´s position plays a critical role, giving very low coupling when the emitter is outside the slot region. These requirements make complex both the fabrication and the emitter integration. Achieving slot widths below 10 nm is beyond the state of the art of almost any fabrication technology and deterministic deposition of a QD with that accuracy can be complicated. To reduce those limitations, we need to optimize the geometry of the cavity further. We have performed a hybrid GA-NN optimization of the Bragg corrugation geometry. The GA-NN optimization must deal with the trade-off between reducing the cavity modal volume (to increase $g$) and maintaining the appropriate $Q$ to achieve $I$ >0.9 with $\gamma^* = 10^4\gamma$. With this aim, we set $\omega_s$ =20 nm and the number of periods to #p=20. The structure without optimization has a modal volume of about $10^{-2}(\lambda/2n)^3$, which gives $I$ =0.82 with $\gamma^* = 10^4\gamma$. Figure 4a shows the GA-NN optimized geometry. Somehow surprisingly to us, the GA-NN found that it is enough to change the widths of the most external Bragg corrugations, leaving the others unperturbed. This geometry provides the best confinement of the cavity mode in the center of the structure, significantly reducing the modal volume while maintaining the correct $Q$.

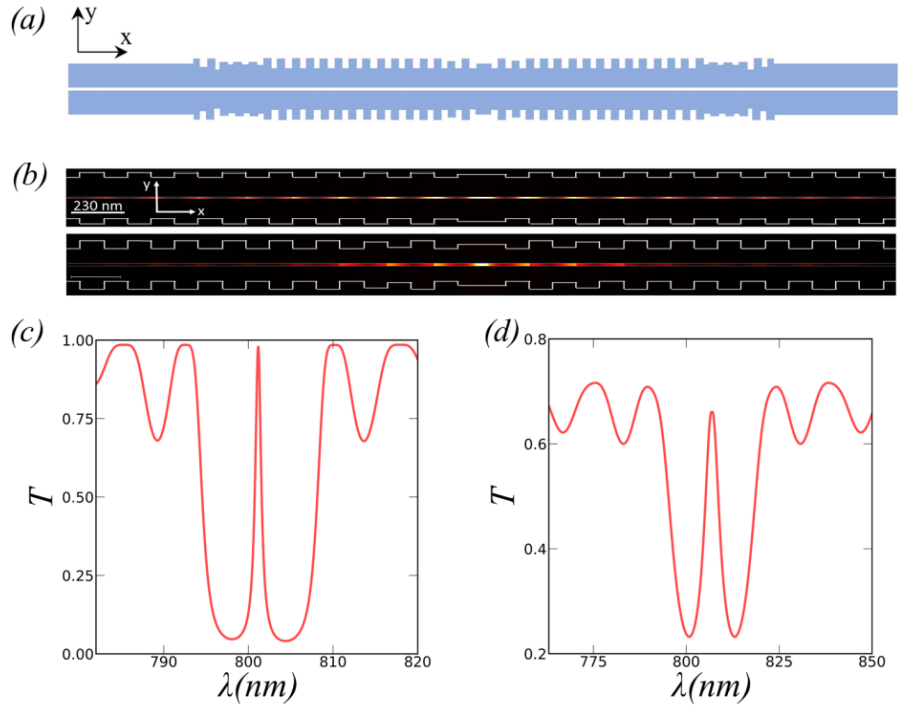

Figure 4. (a) Optimized structure for fixed ($\omega_s$, #p) = (20,20). Cavity-mode field profile in the XY plane inside the cavity region for (b) Cavity mode profiles of the non-optimized structure (up) and optimized structure (down); (c) GA-NN optimized structure. Transmission spectra for (d) Structure without optimization; (e) GA-NN optimized structure.

Figure 4 shows the cavity-mode profile and the transmission spectrum for the structure with and without optimization. It is easy to appreciate how the cavity mode is significantly more confined in the central region of the optimized cavity. The modification of the widths of the external Bragg corrugations creates a tapered section that connects the cavity with the input/output slot waveguides and increases the confinement of the cavity mode. The modal volume is reduced from $7 \cdot 10^{-3}(\lambda/2n)^3$ to $2.5 \cdot 10^{-3}(\lambda/2n)^3$, a factor of 2.8. At the same time, the *FWHM* has been increased to $Q=50$, keeping the system in the region of high *I*. The reduction in the modal volume, and the *Q* adjustment, improve the indistinguishability from *I* =0.82 to *I* =0.91. In conclusion, we obtain that for the optimized structure, we can achieve *I* >0.9 for $\gamma^* = 10^4 \gamma$ with a slot width of $\omega_s=20$ nm, relaxing the tight requirements for the fabrication of the slot to more realistic values. The resulting transmission spectra of the optimized device reveals that there is a 7 nm shift of the resonance wavelength. This results from the discontinuous alteration of the periodicity of the Bragg reflectors. The $\lambda$-Bragg condition for total reflection changes along the corrugations, giving rise to a small modification of the spectra. This resonance displacement could be reduced through a second optimization process involving the maximization of *I* together with the minimization of the $\lambda$-shift, which will be covered in future works.

Although simulations results show a promising device performance, potential difficulties related to fabrication have to be considered according to CMOS compatible processes. Realization of vertical slots widths below 80 nm can be difficult with standard lithography techniques. For emitters with $\gamma^*=10^2 \gamma$ slots between 36 nm and 45 nm are needed (see Table

1). Despite achieving these widths can be challenging, there are many experimental demonstrations reporting the fabrication of sub-100 nm slots (between 30 nm and 80 nm) using e-beam lithography (EBL)[57-64]. On the other hand, strong dissipative emitters with $\gamma^*=10^4\gamma$ require slot widths between 5 nm and 15 nm. Defining sub-10 nm structures with EBL is a great challenge, requiring simultaneous control of several factors like resist contrast, beam diameter, resist development mechanics, and limitations in metrology[65]. A fabrication procedure with EBL is reported in [65], which allows to achieve slots widths down to 8 nm in Si substrates. Also, in [66] the authors experimentally demonstrate a different fabrication approach achieving slots with 10 nm width in Si waveguides. In this context, the relaxation up to 20 nm width achieved through the ML optimization is especially relevant, since it reduces the fabrication requirements from the limit of the technology (5 nm) to a more accessible value (20 nm). Still, we must emphasize that achieving such ultra-narrow slots presents a significant challenge which require top-state-of-the-art resolution technology.

Another key aspect to consider for the experimental realization is the nanoscale positioning approach for the deposition of the QE in the 20 nm slot region of the cavity. Recently, several nanoscale positioning techniques compatible with nanofabrication processes have shown promising results, achieving positioning accuracy at the nanometer level[67]. Atomic force microscopy-based positioning approaches with 30 nm positioning accuracy have been reported with GaAs QDs strongly coupled to a nano-cavity[68]. Confocal micro-photoluminescence techniques also showed 10 nm positioning accuracy with GaAs QDs inside a photonic structure[69]. Bi-chromatic photoluminescence approaches with 5 nm position accuracy was recently achieved through a novel image analysis software implementation in the positioning setup[70]. Also, In Situ lithographic techniques, where the QD position extraction and the nanostructure definition are developed in the same setup have improved position accuracy down to 30 nm[71]. Pick-and-place techniques, which are the most suitable approach for our specific structure, have also shown significant progress[72]. In [73] Si vacancy centers were transferred to AlN waveguides achieving 98% coupling efficiency, the placement mean error was about 38 nm. According to this, for a pick and place deposition, assuming a normal distribution we would have a standard deviation of 38 nm with a target of 20 nm, which leads to 34% probability of successful deposition. Therefore, the positioning accuracy required for our structure lies close to the limit of the technology depending on the positioning approach. An experimental realization of a QE coupling would involve fabricating a large number of devices and looking for suitable candidates one at a time. This approach enables the experimental demonstration of specific physical phenomena for quantum information applications, but is still far from a scalable system to incorporate multiple identical QE for more advanced experiments and applications.

**Conclusions**

We explored a hybrid slot-Bragg nanophotonic cavity for the generation of indistinguishable photons at RT from various quantum emitters through a combination of numerical methods. We obtain the values of the theoretical indistinguishability, efficiency and Purcell enhancement for each configuration (i.e. waveguide width, slot width, number of periods). We obtained theoretical near-unity indistinguishability and high efficiency simultaneously by parameter sweep optimization. To relax the fabrication requirements (slot width) for near-unity indistinguishability, we have developed a machine learning algorithm that provides the optimal geometry of the cavity. According to our simulations, the optimized structure shows high indistinguishability ($I$>0.9) with slot widths about 20 nm. The geometrical features of the optimized design present significant challenges from the perspective of fabrication process. Although the device may be far from a real scalable technology it can be suitable for experimental demonstration of single photon operation. Also, the developed ML approach may provide insights for the optimization of different photonic structures for quantum information applications.

*Acknowledgments*

We gratefully acknowledge financial support from the European Union's Horizon 2020 research and innovation program under grant agreement No. 820423 (S2QUIP) and CSIC Quantum Technology Platform PT-001. JML acknowledges the Agencia Estatal de Investigación (AEI) grant PID2019-106088RB-C31.

*Disclosures*

The authors declare no conflicts of interest.

*Data availability statement*

Details underlying the results presented in this paper are available in the Supplement 1 file.

*Funding sources*

This work has been funded by the European Union's Horizon 2020 research and innovation program under grant agreement No. 820423 (S2QUIP).

*Supplement information*

This material is available free of charge via the internet at http://pubs.acs.org. Details of the computation of the indistinguishability; details of the numerical simulation; details of the machine learning algorithm; Information about the fabrication process.

*For table of contents use only*

Title: Numerical optimization of a nanophotonic cavity by machine learning for near-unity photon indistinguishability at room temperature.

Authors: Guimbao Joaquin, Sanchis Lorenzo, Weituschat Lukas, Llorens Montolio Jose, Song Meiting, Cardenas Jaime, Postigo Pablo.

Synopsis: The graphic shows a renderized image of the nanophotonic cavity explored in the paper. A 2D material-based quantum emitter is placed on top of the nanocavity with a laser beam illuminating the sample. At the left of the cavity there is an schematic of a Neural Network representing the machine learning scheme used for the optimization of the geometry of the cavity.

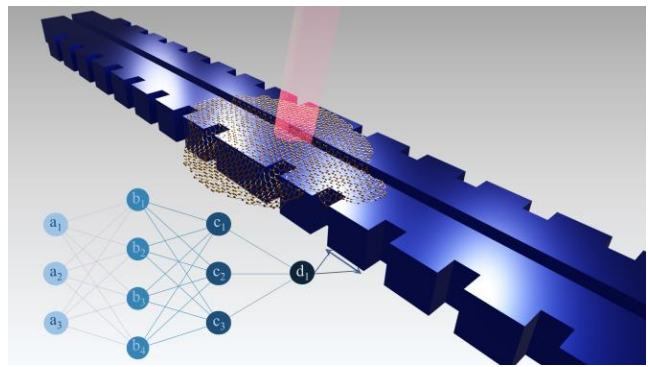